\documentclass[conference]{IEEEtran}
\IEEEoverridecommandlockouts

\usepackage{cite}
\usepackage{amsmath,amssymb,amsfonts}
\usepackage{algorithmic}
\usepackage{graphicx}
\usepackage{textcomp}
\usepackage{xcolor}
\usepackage{url}
\usepackage{hyperref}
\def\BibTeX{{\rm B\kern-.05em{\sc i\kern-.025em b}\kern-.08em
    T\kern-.1667em\lower.7ex\hbox{E}\kern-.125emX}}

\makeatletter
\newcommand{\linebreakand}{%
  \end{@IEEEauthorhalign}
  \hfill\mbox{}\par
  \mbox{}\hfill\begin{@IEEEauthorhalign}
}
\makeatother    
\begin{document}

\title{AI Engineering Blueprint for On-Premises Retrieval-Augmented Generation Systems
}

\author{\IEEEauthorblockN{1\textsuperscript{st} Nicolas Weeger}
\IEEEauthorblockA{\textit{Ansbach UAS} \\
Ansbach, Germany \\
nicolas.weeger@hs-ansbach.de}
\and
\IEEEauthorblockN{2\textsuperscript{nd} Jakob Winkler}
\IEEEauthorblockA{\textit{University of Gießen} \\
Gießen, Germany \\
jakob.m.winkler@uni-giessen.de}
\and
\IEEEauthorblockN{3\textsuperscript{rd} Annika Stiehl}
\IEEEauthorblockA{\textit{Ansbach UAS} \\
Ansbach, Germany \\
annika.stiehl@hs-ansbach.de}
\linebreakand
\IEEEauthorblockN{4\textsuperscript{th} Jóakim von Kistowski}
\IEEEauthorblockA{\textit{Aschaffenburg UAS} \\
Aschaffenburg, Germany \\
joakim.vonkistowski@th-ab.de}
\and
\IEEEauthorblockN{5\textsuperscript{th} Christian Uhl}
\IEEEauthorblockA{\textit{Ansbach UAS} \\
Ansbach, Germany \\
christian.uhl@hs-ansbach.de}
\and
\IEEEauthorblockN{6\textsuperscript{th} Stefan Geißelsöder}
\IEEEauthorblockA{\textit{Ansbach UAS} \\
Ansbach, Germany \\
stefan.geisselsoeder@hs-ansbach.de}
}

\maketitle

\begin{abstract}
Retrieval-augmented generation (RAG) systems are gaining traction in enterprise settings, yet stringent data protection regulations prevent many organizations from using cloud-based services, necessitating on-premises deployments. While existing blueprints and reference architectures focus on cloud deployments and lack enterprise-grade components, comprehensive on-premises implementation frameworks remain scarce.

This paper aims to address this gap by presenting a comprehensive AI engineering blueprint for scalable on-premises enterprise RAG solutions. It is designed to address common challenges and streamline the integration of RAG into existing enterprise infrastructure. The blueprint provides: (1) an end-to-end reference architecture described using the 4+1 view model, (2) a reference application for on-premises deployment, and (3) best practices for tooling, development, and CI/CD pipelines, all publicly available on GitHub\footnote{\url{https://github.com/aiengineeringblueprints/Enterprise_RAG_Blueprint}}. Ongoing case studies and expert interviews with industry partners will assess its practical benefits.

\end{abstract}


\section{Introduction}
Retrieval-Augmented Generation (RAG) has emerged as a powerful technique for enhancing the capabilities of Large Language Models (LLMs) by integrating external knowledge sources~\cite{lewisRetrievalAugmentedGenerationKnowledgeIntensive2020}. 

In addition to the internal regulations and safety issues of companies~\cite{brehmeRetrievalAugmentedGenerationIndustry2025a}, stringent data protection regulations, such as the EU AI Act, the GDPR in Europe, and HIPAA in the healthcare sector, restrict the use of cloud-based LLM services for processing personal or sensitive data, as it involves transferring data outside of the company's IT infrastructure~\cite{wadaRetrievalaugmentedGenerationElevates2025, he}. Research initiatives, including the EU's OpenEuroLLM~\cite{OpenEuroLLMEuropeanFamily2025} and the Helmholtz Foundation Model Initiative~\cite{HelmholtzFoundationModel2025}, highlight the importance of LLMs that comply with these laws. Consequently, many companies opt for on-premises solutions to maintain oversight of their data and adhere to strict data security and compliance standards.

Building and deploying RAG systems at an enterprise level poses significant challenges. A recent MIT study from 2025~\cite{challapallyGenAIDivideState2025} highlights this disparity, showing that, while custom AI tools are widely investigated in enterprises, only 5\% are successfully deployed in production. A primary reason for this low success rate is the complexity that arises when scaling these systems and integrating them into existing enterprise infrastructure. Several studies emphasize the challenges of building and deploying RAG systems, particularly in enterprise settings~\cite{borahApplicationsArtificialIntelligence2022,schonbergerARTIFICIALINTELLIGENCESMALL2023,bruckhausRAGDoesNot2024}. These challenges include the need for specialized expertise in AI and data management, the complexity of integrating RAG systems with existing IT infrastructure, and the difficulty of ensuring data security and compliance with regulations.

This research aims to address the challenges associated with building and deploying on-premises RAG systems by providing a simple yet comprehensive, end-to-end blueprint. The main contributions of this paper are:
\begin{itemize}
\item An end-to-end reference architecture for on-premises enterprise RAG, formally described using the 4+1 view model, with explicit architectural trade-offs and variation points.
\item A deployable reference application that serves as an adaptable starting point for integrating RAG into existing enterprise infrastructure, including all enterprise-grade components.
\item Best practices for tooling, development, and CI/CD pipelines tailored to on-premises container deployment.
\end{itemize}

\section{Related Work}

RAG systems represent an emerging area of research, with a growing body of literature exploring various aspects of RAG, including its theoretical foundations~\cite{lewisRetrievalAugmentedGenerationKnowledgeIntensive2020}, agentic RAG systems~\cite{singhAgenticRetrievalAugmentedGeneration2025,nguyenMARAGMultiAgentRetrievalAugmented2025}, and other methods to improve the performance of the RAG system outputs~\cite{liEnhancingRetrievalAugmentedGeneration2025a}. The literature also covers evaluation metrics and frameworks~\cite{esRAGAsAutomatedEvaluation2024,yangHotpotQADatasetDiverse2018a,yuEvaluationRetrievalAugmentedGeneration2025a} or the security of RAG systems~\cite{ammannSecuringRAGRisk2025a,zengMitigatingPrivacyIssues2025a}.

The concept of RAGOps~\cite{xuRAGOpsOperatingManaging2025} has emerged from \mbox{LLMOps}~\cite{diaz-de-arcayaLargeLanguageModel2024} by integrating data lifecycle management as a core component. This paradigm characterizes the architecture of RAG applications and outlines the complete lifecycle of these composite systems using the 4+1 architectural view model. RAGOps defines design considerations and provides quality tradeoffs for the various stages of the RAG system lifecycle. 

In order to understand industry adoption, Brehme et al.~\cite{brehmeRetrievalAugmentedGenerationIndustry2025a} reveal key insights from 13 practitioners. The study indicates that contemporary RAG applications are frequently domain-specific question-answering systems that are predominantly in the prototype stage. The industry's prevailing requirements place significant emphasis on data protection, security, and quality. Nevertheless, the preprocessing of data remains a considerable challenge, with evaluation being primarily conducted by human analysts. 

Since practical implementation guidance for RAG systems has not yet been deeply explored in peer-reviewed publications, grey literature provides valuable supplementary insights. Blog posts and white papers discuss, for example, common challenges~\cite{changRetrievalaugmentedGenerationRealized2024}, specific RAG architectures~\cite{kelly8RetrievalAugmented2025}, and implementation patterns and anti-patterns~\cite{fowlerEmergingPatternsBuilding2025}. However, the majority of these discussions focus on cloud-based RAG systems~\cite{vanteylingenBuildingEnterpriseRAG2025}, while only few address on-premises deployments~\cite{DesigningOnpremisesArchitecture2025}.

A number of open-source RAG implementations are available on GitHub, including RAGFlow\footnote{RAGFlow, \url{https://github.com/infiniflow/ragflow}, accessed: Feb. 9, 2026}, kotaemon\footnote{kotaemon, \url{https://github.com/Cinnamon/kotaemon}, accessed: Feb. 9, 2026}, and the FELDM RAG Blueprint\footnote{RAG Blueprint, \url{https://github.com/feld-m/rag_blueprint}, accessed: Feb. 9, 2026}. However, these systems often lack the components necessary for enterprise-level scaleability.

Despite this broad theoretical foundation, comprehensive end-to-end reference architectures with code-level implementation guidance remain scarce~\cite{weegerPracticableMachineLearning2025} — particularly for on-premises deployments, where data security and regulatory compliance are critical concerns.

\section{Proposed Blueprint}

This blueprint developed in this study builds on the conceptual RAGOps framework~\cite{xuRAGOpsOperatingManaging2025} and translates it into a concrete, deployable architecture for on-premises environments. While RAGOps provides a paradigm for operating RAG systems, it remains at the conceptual level without offering implementation artifacts. Existing open-source RAG implementations provide functional applications but lack enterprise-grade components, such as access control, guardrails, and observability, and do not provide structured architectural documentation to support adaptation to existing enterprise infrastructure. This blueprint addresses both of these gaps by combining a formally described reference architecture with a deployable reference application and CI/CD pipelines.

The architectural design follows the 4+1 view model by Kruchten~\cite{kruchten4+1ViewModel1995}. Detailed views and components can be found in the provided GitHub repository. The accompanying reference application implements all architectural components as a generic, adaptable starting point for enterprise integration, including documentation and a CI/CD pipeline for on-premises container deployment.

The blueprint focuses on the RAG application layer. The deployment of LLMs and embedding models, while a pivotal prerequisite, is regarded as a distinct architectural component addressed in separate research~\cite{diaz-de-arcayaLargeLanguageModel2024}.

\subsection{Requirements}
The architectural design is informed by the requirements identified in the extant literature. Bruckhaus~\cite{bruckhausRAGDoesNot2024} emphasizes data protection and security, accuracy and explainability, as well as seamless enterprise integration and scalability. Xu et al.~\cite{xuRAGOpsOperatingManaging2025} focus on operational aspects, highlighting monitoring and observability, adaptability, traceability, and reliability. Brehme et al.~\cite{brehmeRetrievalAugmentedGenerationIndustry2025a} confirm and extend these findings through interviews with 13 industry practitioners, identifying additional requirements such as continuous operation and explainability. We adopt the requirements identified by Brehme et al. as our primary basis. However, some requirements, such as usability and costs, are excluded as they depend on specific deployment scenarios. The following requirements are considered:

\begin{itemize}
    \item Security and Data Protection
    \item Quality, Relevance and Accuracy
    \item Explainability and Transparency
    \item Performance
    \item Continuous Learning
    \item Continuous Operation
    \item Integration in Setup
    \item Scalability
    \item Licensing and Copyright
    \item Ethical Considerations and Bias
\end{itemize}

\subsection{Architecture}

\begin{figure}[b]
\centerline{\includegraphics[width=\linewidth]{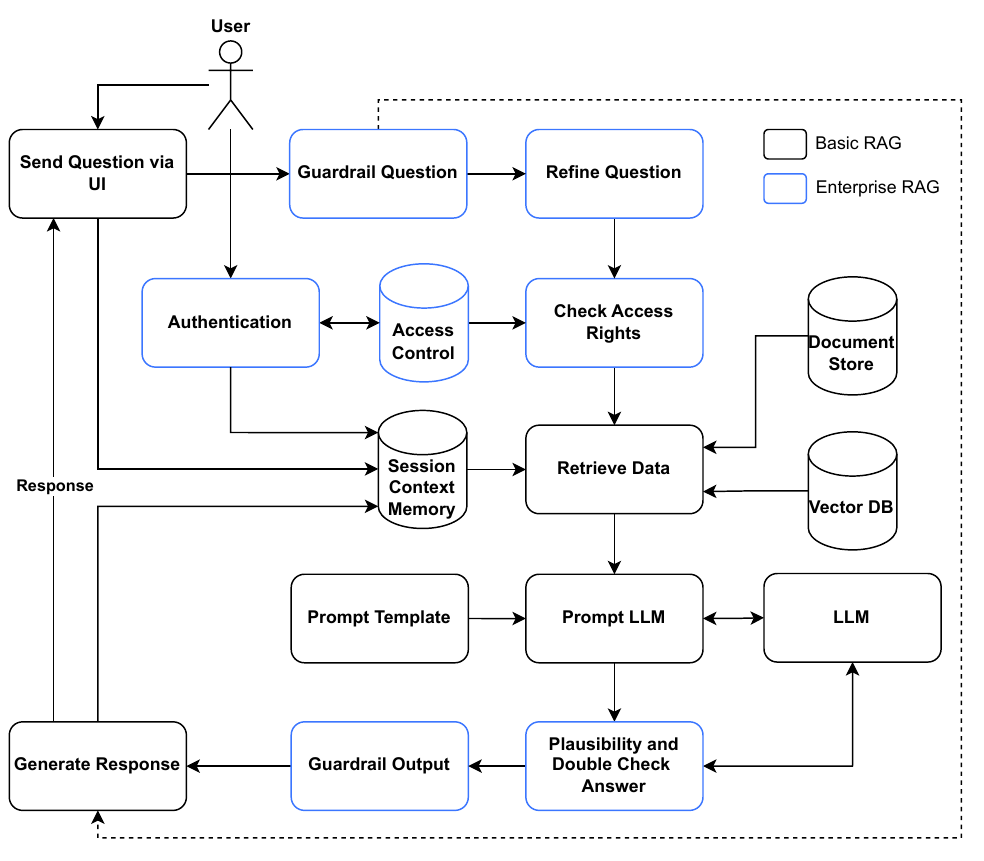}}
\caption{Proposed functional architecture for enterprise RAG.}
\label{fig:functional_view}
\end{figure}

Figure~\ref{fig:functional_view} shows the functional architecture, which is designed to meet these requirements and address the challenges associated with building and deploying enterprise ready RAG systems. The functional architecture is structured in two stages: a basic RAG stage (black boxes) providing core retrieval and generation functionality, and an enterprise stage (blue boxes) adding components required for production use in enterprise contexts, such as access control, guardrails, and monitoring. This staged design allows organizations to start with a basic RAG deployment and incrementally adopt enterprise components based on their specific requirements.

The authentication and access control components are designed to ensure security and data protection in enterprise scenarios involving different document access rights. Thus, the retrieved data contains only the information which the user is authorized to access. The guardrails, query refinement, and answer verification components employ additional LLM calls to filter inappropriate requests and mitigate bias, reformulate user queries for improved retrieval, and validate generated responses, thereby addressing the requirements of quality, relevance, accuracy, and ethical considerations. While these additional LLM calls increase response latency, they are essential for meeting enterprise quality and compliance requirements.

\begin{figure}[b]
\centerline{\includegraphics[width=\linewidth]{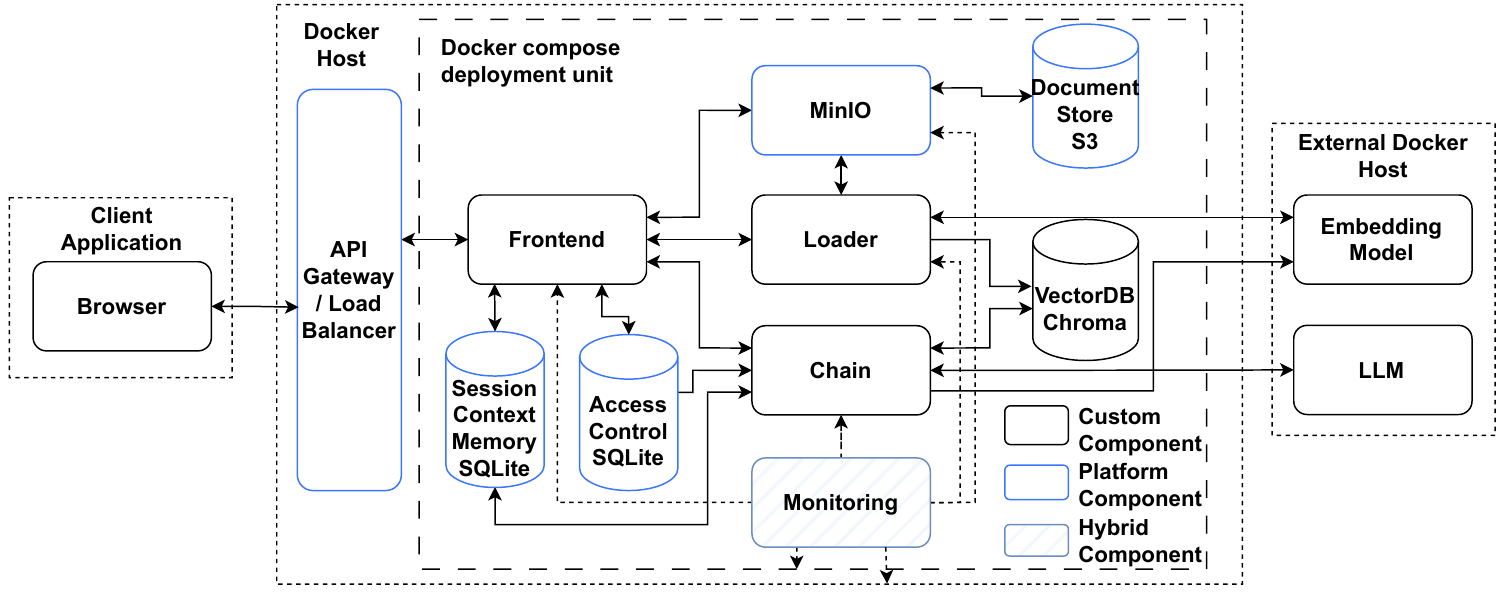}}
\caption{On-premises deployment of proposed RAG components.}
\label{fig:deployment_view}
\end{figure}

Figure~\ref{fig:deployment_view} shows the deployment of on-premises RAG components. In contrast to the functional view, blue boxes here denote platform modules that serve as substitutable placeholders for existing enterprise infrastructure. This deployment is designed to address the requirements of integration in setup and scalability. Rather than making assumptions about specific infrastructures, we provide a solution that is easily adaptable using Docker Compose, enabling quick deployment trials. The components are designed as microservices, separated by RESTful APIs. This design trades minimal network latency overhead for the ability to scale and replace components independently, in order to meet performance requirements.

The platform modules, denoted by blue boxes, such as the S3 storage and databases, are designed as `dummy' components to provide the necessary interfaces. These modules can be readily substituted with existing enterprise infrastructure, for instance, replacing the bundled MinIO storage with an existing S3-compatible enterprise storage solution.

When combined with locally deployed LLMs and embedding models, the architecture ensures that all data processing remains within the enterprise's IT infrastructure. This directly addresses the requirements relating to security, data protection, licensing and copyright.

The Frontend component has been designed as a test user interface, as it will need to be adapted for each company based on their technologies and design preferences. To enhance explainability, the architecture provides interfaces to display the sources and used chunks for answer generation. Nevertheless, explainability and transparency are important requirements that could be further enhanced through additional design and implementation efforts. 

The Chain component is responsible for retrieving relevant information from the knowledge base based on the user's query. This part is responsible for quality, relevance, and accuracy. Enterprises may choose to extend this component when specifically focusing on improving these requirements. 

The Loader component provides the interface for loading data into the system. It is designed to be flexible and adaptable to different data sources and formats, allowing enterprises to easily integrate their existing data into the RAG system.

The Monitoring component is responsible for observability and continuous operation. It utilizes OpenTelemetry\footnote{OpenTelemetry, \url{https://opentelemetry.io}, accessed: Feb. 11, 2026} to collect telemetry data across all three pillars, namely logs, metrics, and traces, ensuring the system's performance and health. This setup allows for easy integration with various observability stacks, whether local or cloud-based.

\section{Summary and Future Work}

This paper presents an AI engineering blueprint that combines a formally described reference architecture with a deployable reference application and CI/CD pipelines for on-premises enterprise RAG systems.

It is in the early stages of development. The architecture design and reference application have completed the first implementation phase and are available for further evaluation and refinement. 

Industry partners are currently being engaged to evaluate the reference application and architecture in real-world contexts. The evaluation will be conducted according to Design Science Research (DSR) methodology~\cite{hevnerDesignScienceInformation2004}, and will include quantitative and qualitative methods such as expert interviews~\cite{glaeserExperteninterviewsUndQualitative2010} and case studies~\cite{runesonGuidelinesConductingReporting2009}. The results of the evaluation will be used to refine and improve the blueprint based on the feedback from industry partners and the evaluation results. 

The following potential improvements have been identified:
\begin{itemize}
\item Integration of more advanced RAG approaches, such as Agentic RAG, multimodality or reranking strategies
\item Evaluation metrics and frameworks for retrieval performance, generation quality, and system performance
\item Components to support continuous learning and improvement, explainability and transparency
\end{itemize}

The blueprint aims to lower the barrier for enterprises to adopt on-premises RAG in production.

\bibliographystyle{IEEEtran}
\bibliography{IEEEabrv,My_Library}

\end{document}